\documentclass[preprint, nofootinbib, aps]{revtex4}

\usepackage{hyperref}

\def\[{\left\lbrack}
\def\]{\right\rbrack}

\def\({\left(}
\def\){\right)}
\newcommand{\be}{\begin{equation}}
\newcommand{\ee}{\end{equation}}
\newcommand{\ea}{\end{eqnarray}}
\newcommand{\ba}{\begin{eqnarray}}
\newcommand{\la}{\langle}
\newcommand{\ra}{\rangle}
\newcommand{\asp}{\textquotedblleft}  

\newcommand{\ep}{{\epsilon}}

\newcommand{\cl}{{\cal L}}

\newcommand{\cd}{{\cal D}}

\newcommand{\fdu}{{^{\star}F}}

\newcommand{\jdu}{{^{\star}J}}
\newcommand{\ddu}{{^{\star}d}}
\newcommand{\tdu}{{^{\star}\theta}}

\newcommand{\fmnd}{F_{\mu \nu}}

\newcommand{\prt}{{\partial}}
\newcommand{\diag}{\mbox{diag}}
\newcommand{\sdot}{\!  \cdot \!}
\newcommand{\lra}{\leftrightarrow}
\newcommand{\tr}{\mbox{tr}}
\newcommand{\tht}{\tilde \theta}
\newcommand{\hepth} {hep-th/}

\begin{document}

\title{Issues on 3D Noncommutative Electromagnetic Duality}

\author{Davi C. Rodrigues and Cl\'ovis Wotzasek}
\email{cabral, clovis@if.ufrj.br}
\affiliation{Instituto de F\'{\i}sica, Universidade Federal do Rio de Janeiro,\\
21945-970, Rio de Janeiro, RJ, Brasil}


\begin{abstract}
We extend the ordinary 3D electromagnetic duality to the noncommutative (NC) space-time through a Seiberg-Witten map  to second order in the noncommutativity parameter $\theta$, defining a new scalar field model. There are similarities with the 4D NC duality, these are exploited to clarify properties of both cases. Up to second order in $\theta$, we find that duality interchanges the 2-form $\theta$ with its 1-form Hodge dual ${^\star}  \theta $ times the gauge coupling constant, i.e.,  $ \theta \rightarrow {^\star}  \theta \; g^2$ (similar to the 4D NC electromagnetic duality). We directly prove that this property is false in the third order expansion in both 3D and 4D space-times, unless the slowly varying fields limit is imposed. Outside this limit, starting from the third order expansion, $\theta$ cannot be rescaled to attain an S-duality. In addition to possible applications on effective models, the 3D space-time is useful for studying general properties of NC theories. In particular, in this dimension, we deduce an expression that significantly simplifies the Seiberg-Witten mapped Lagrangian to all orders in $\theta$.


\end{abstract}

\maketitle

\section{Introduction}

	The 4D noncommutative (NC) electromagnetic duality, up to the subleading order in $\theta$ or in the slowly varying fields limit \cite{sw, esw}, via the Seiberg-Witten map \cite{sw}, relates two $U(1)$ gauge theories and has a curious property \cite{ganor, aschieri, 4D}: being $\theta$ the noncommutativity parameter of one of them, the other has ${^\star} \theta g^2$ as its noncommutativity parameter (where $^\star$ is the Hodge duality operator and $g^2$ is the gauge coupling constant). This is more than a simple curiosity, it suggests a consistence problem \cite{ganor, aschieri}. Employing the standard quantization programme, it is well known that a time-like noncommutativity parameter ($\theta^{\mu \nu} \theta_{\mu \nu} < 0$) leads to unitarity violation \cite{uni}. Since $\theta$ is space-like iff $^\star \theta$ is time-like, the above results suggest that a modification on the quantization programme of NC theories is necessary  \cite{altqua}, otherwise only light-like noncommutativity $(\theta^{\mu \nu} \theta_{\mu \nu} = 0 )$ may be consistent with the $U(1)$ NC theory \cite{aschieri}\footnote{It should be noted that in this work we are concerned with the issue of duality of NC theories within the field theoretical framework. From the string theory perspective, $S$-duality of $IIB$ strings in the presence of a magnetic background induces a duality between spacially NC Yang-Mills $\cal N$$=4$ theory  with a string model called NCOS (noncommutative open string), as conjectured in Ref. \cite{stringd}. Although the approaches of Ref. \cite{stringd} and ours are quite different, there are similarities in the resulting dualities, like the exchange of $\theta$ with $ {^\star} \theta g^2$. See our Conclusions for further comments.}.

Being the role of electromagnetic duality in NC theories so relevant, in this work we extend it to the 3D space-time and evaluate the necessity of the slowly varying fields limit from a classical field theoretical perspective, in order to find what are the  fundamental properties of this duality. Many arguments of Ref. \cite{ganor} depend on the space-time dimension (e.g., the 4D space-time is the only one which $\theta$ and $^\star \theta$ are both 2-forms and the $S$-dual massless gauge fields are both 1-forms), therefore a natural question is how the NC electromagnetic duality presents itself in other dimensions, and to what extent the properties of the 4D NC electromagnetic duality can be extended to those. From all possibilities, the 3D space-time seems to be a natural option. In this space-time, we establish to second order in $\theta$ the dual scalar action (consistently with the rule $\theta \rightarrow \, ^\star \theta g^2$) and we show that many terms of the Seiberg-Witten mapped action can be considerably simplified. The necessity of the slowly varying fields limit, to preserve the rule $\theta \rightarrow \, ^\star \theta g^2$ and therefore $S$-duality\footnote{In the sense of a global inversion of the coupling constant (string $S$-duality is not of concern in this approach).}, starts from the third order in $\theta$ for any space-time dimension (with $D \ge 2$).


This Paper is organized as follows: after a review of the ordinary 3D electromagnetic duality, we establish its extension to the NC space-time up to first order in $\theta$, providing the duality map and some physical details. In the fourth section, we extend this duality to second order and, through the Seiberg-Witten differential equation, analyse its behaviour to higher orders. Finally, in the last section, we present our conclusions.

\vspace{.5in}
\section{Revisiting the 3D electromagnetic duality}

To introduce our framework, we briefly review the electromagnetic duality in 3D ordinary space-time. The electromagnetic theory action with a 1-form source $J$ is
\be
	\label{e1}
	S_A[A,J] = \int \( a \; F \wedge \fdu \; + \; e \; A \wedge \jdu \),
\ee
where $A$ is the 1-form potential, the field strength $F$ satisfies, by definition, $F = dA$ and $a = -1/(2g^2)$. To preserve gauge invariance and to satisfy the continuity equation, $\jdu$ must be a closed 2-form.

As usual, the dynamics of the electromagnetic fields comes from the equation of motion and the Bianchi identity, namely,
\be
	\label{em1}
	d \fdu = - \frac e {2a} \; \jdu \;\;\;\;\;\;\; \mbox{and} \;\;\;\;\;\;\; dF = 0.
\ee 
Except for the sign of the first equality, the above equations are valid in any space-time dimension. These equations can be expressed on more \asp observational grounds" through the electric and magnetic fields given by $E^i = F^{i0}$ and $B= \vec \nabla \times \vec A = - \ep^{ij}\prt_i A_j = - F_{12}$. We adopt the conventions $g = \diag (+, -, -)$, $F = \frac 12 F_{\mu \nu} dx^\mu \wedge dx^\nu$, $\vec E = (E^1, E^2)$, $\vec \nabla = (\prt_1, \prt_2)$, $\ep_{12} = \ep^{12} = 1$,  greek indices can assume the values 0, 1 or 2 and latin indices only 1 or 2. Using this notation, Eqs. (\ref{em1}) becomes
\ba
	\label{dive}
	&&\vec \nabla \cdot \vec E  =  \frac e {2a} \rho, \\[.1in]
	\label{rote}
	&&\vec \nabla \times B = \dot {\vec E} + \frac e {2a}  \vec j , \\[.1in]
	\label{rotb}
	&&\vec \nabla \times \vec E = - \dot B,  
\ea
where $J^\mu = (\rho, \vec j)$, a dot over a field means temporal derivative and, by definition, $(\vec \nabla \times B)^i = \ep^{ij} \, \prt_j B$. These equations have curious similarities and differences with usual 4D Maxwell equations. Among the differences, since $\vec E$ is a vector and $B$ a pseudo-scalar, even in the case without sources, there is no hope of finding a simple duality which simply interchanges electric and magnetic fields. However, Eq.(\ref{dive}) with $\rho = 0$ hints to set $\vec E = - \vec \nabla \times \phi$, which implies $F^{i0} = - \ep^{ij} \prt_j \phi$. Thus, to preserve Lorentz symmetry, we shall set
\be
	\label{map1}
	F^{\mu \nu} = \ep^{\mu \nu \lambda} \prt_\lambda \phi \;\;\;\;\;\; \mbox{(for $J = 0$)}.
\ee	
Consequently, $B = - \dot \phi$. Using the map (\ref{map1}), it is straightforward to show that Eqs.(\ref{dive}) and (\ref{rote}), without sources, turn into an identity, while (\ref{rotb}) becomes the free scalar field equation,	$\prt_\mu \prt^\mu \phi = 0$. Note that there is no violation of the number of degrees of freedom, both descriptions (vectorial and scalar) have one degree of freedom.

To conclude this introduction, we shall present the 3D electromagnetic duality with a source $J$ and introduce the master Lagrangian approach \cite{master}. Consider the action
\be
	\label{m1}
	S_{M}[F, \phi] = \int \[ a F \wedge \( \fdu + \frac {e} a \Lambda \) -  d \phi \wedge F \],
\ee
where $F$ is regarded as an independent 2-form and $\Lambda$ is a 1-form. Equating to zero the variation of the above action with respect to $\phi$,  we obtain $dF =0$. This implies, in Minkowski space, that $F=dA$. Replacing $F$ by $dA$ and setting $\jdu = d \Lambda$, $S_M$ becomes equivalent to the action in Eq. (\ref{e1}). 

On the other hand, the variation of Eq.(\ref{m1}) with respect to $F$ produces
\be
	\label{map2}
	\fdu =  \frac 1 {2a} \( d\phi - e \Lambda \).
\ee 

Inserting Eq.(\ref{map2}) into the master action $S_M$ (and recalling ${^\star}{^\star} =1$ for any differential form in the dealt space-time), we find

\be
	\label{sp}
	S_{M} [F, \phi] \leftrightarrow - \frac 1 {4a} \int (d\phi - e\Lambda) \wedge {^\star}(d\phi - e\Lambda) 	= S_{\phi_\Lambda} [\phi].  
\ee
	
We use the symbol \asp $\leftrightarrow$" instead of \asp $=$" to be clear that equivalence of actions (functionals) is to be understood as a correspondence between their equations of motion; that is, if $ S_1 \leftrightarrow S_2$, the set of equations of $S_1$ can be manipulated, using its own equalities, or inserting new redundant ones, to become the set of equations of $S_2$ (the inverse also proceeds). 

The two equations of motion of $S_M$ ($dF = 0$ and Eq.(\ref{map2})) generate a map between the equations of motion of $S_A$ and $S_\phi$, viz,
\be
	{^\star dA} =  \frac 1 {2a} \( d\phi - e \Lambda \).
\ee 
Applying $d$ on both sides, we find Eq.(\ref{em1}), while the application of $d^\star$ results in  $d^\star d \phi =  e \, d^\star \Lambda$, which is the equation of motion of $S_\phi$.

\vspace{.5in}

\section{3D NC  electromagnetic duality to first order in $\theta$}

The NC version of the $U(1)$ gauge theory, whose gauge group we denote by $U_*(1)$, is given by \cite{rev}
\be
	\label{sm}
	S_{\hat A*} = a \int \hat F \wedge_*  {^\star} \hat F ,
\ee
where $a$ is a constant, $\hat F = d \hat A - i \hat A \wedge_* \hat A = \frac 12 (\prt_\mu \hat A_\nu - \prt_\nu \hat A_\mu - i[\hat A_\mu, \hat A_\nu]_* ) d x^\mu \wedge dx^\nu$, $[A,B]_* = A*B - B*A$ and
\be
	(A * B)(x) = \exp \( \frac i 2 \theta^{\mu \nu} \prt^x_\mu \prt^y_\nu \) A(x) \; B(y)|_{y \rightarrow x}
\ee
is the Moyal product. In particular, $[x^\mu, x^\nu]_* = i \theta^{\mu \nu}$. $(\theta^{\mu \nu})$ can be any real and constant anti-symmetric matrix.

Since $d \hat F \not= 0$, previous duality arguments cannot be directly applied. However, Seiberg and Witten have shown, for infinitesimal gauge transformations, that a $U_*(N)$ gauge theory can be mapped into a $U(N)$ one \cite{sw}. As a corollary, also useful to our purposes, this map provides a more direct treatment of the observables \cite{jackiw}. To first order in $\theta$, for the $U(1)$ case, this map reads,
\be
	\label{sw}
	\hat A = \[ A_\mu - \theta^{\alpha \nu} A_\alpha \( \prt_\nu A_\mu - \frac 12 \prt_\mu A_\nu \) \] dx^\mu
\ee
in which $\hat A$ transforms as $\delta_{\hat \lambda} \hat A = d \hat \lambda - 2 i \hat A \wedge_* \hat \lambda$, while $A$ as $\delta_\lambda A = d \lambda$. 

Inserting (\ref{sw}) into action (\ref{sm}) one obtains an effective $U(1)$ gauge theory whose action contains explicit $\theta$ corrections. The action for this effective theory is denoted by $S_{A_\theta}$ and the products between its fields are the ordinary ones. Up to first order in $\theta$, 
\ba
	\label{a*}
	S_{A_\theta} 	&& =  a \int F \wedge \fdu \; \( 1 + \la \theta, F \ra \)  \\[.1in]
	&& = - a  \int  \( \vec E^2 - B^2 \) \(1 - \vec \theta \sdot \vec E - \theta B \) d^3x, \nonumber
\ea
where $F=dA$, $(\vec \theta)^i = \theta^{i0}$, $\theta = \theta^{12}$ and $ \la \; , \, \ra$ is the scalar product between differential forms\footnote{In odd dimensional Minkowski space, the internal product of two n-forms $A$ and $B$ is defined by $\la A, B \ra = {^\star ( {^\star A} \wedge B)} = \frac 1{n!} A_{\mu_1 \; ... \; \mu_n} B^{\mu_1 \; ... \; \mu_n}$.}, in particular $\la F, \theta \ra = {^\star}( \fdu \wedge \theta) = \frac 12 \theta^{\mu \nu} F_{\mu \nu}$. In order to $S_{A*}$ be dimensionless, the constant $a$ must have dimension of length. The term $F^{\mu \nu} F_{\nu \lambda} \theta^{\lambda \kappa} F_{ \kappa \mu} \, d^3x$ which appears in 4D electromagnetism, also occurs in 3D, but it is proportional to $\fdu \wedge F \,  \la F, \theta \ra$ \cite{ghosh}. The equations of motion are
\be
	\vec \nabla \cdot  \vec D  = 0,
\ee
\be
	\vec \nabla \times H = \dot{\vec D},
\ee
\be
	\label{eqt3}
	\vec \nabla \times \vec E = - \dot B.
\ee
In above, $\vec D = \vec E (1 - \vec \theta \cdot \vec E - \theta B) - \frac 12 \vec \theta (\vec E^2 - B^2)$ and $H = B (1 - \vec \theta \cdot \vec E - \theta B) + \frac 12 \theta (\vec E^2 - B^2)$ (these definitions are analogous to the one used in Ref.\cite{testing}). Eqs. (\ref{rotb}) and (\ref{eqt3}) are equal because both comes from the Bianchi identity. Clearly $\vec \theta$ is responsible for a violation of spacial isotropy.

Exploiting the Bianchi identity, we propose the following master action:
\be
	\label{amt}
	S_{M_\theta}[F, \phi] = \int \[ a \, \fdu \wedge F \, \( 1 + \langle \theta, F \rangle \)  - d \phi \wedge F \].
\ee
We will use the above master action to find the first order duality, and a natural generalization of it will be employed to unveil the duality in higher $\theta$ orders. 

However, this is not the only possible master action, the following actions also ascertain dualities between the same vector and scalar descriptions of NC 3D electromagnetism:
\be
	S_{M_{\theta,c}} [G, \phi] = \int \[ a \; G \wedge {^\star G} \; ( 1 + c \la \theta, G \ra ) - \( 1 + \frac 12 (c-1) \la \theta , G \ra \) d \phi \wedge G \],
\ee

\be
	\label{aml}
	S_{M'_\theta}[B, A] =  \int \[ - \frac 1{4a} B \wedge {^\star B} \( 1 - \frac 1{2a} \la \theta, {^\star B} \ra \)  - B \wedge dA \].
\ee

The first one is a generalization of the master action in Eq.(\ref{amt}) by a continuous and arbitrary parameter $c$, being the latter recovered for $c=1$. The master $S_{M_{\theta,c}}$ has the interesting feature of balancing the NC contribution between its two terms. Nevertheless, for any $c$, the models it connects are the same vector and scalar ones that are found by $S_{M_{\theta}}$. In Eq.(\ref{aml}), $A$ and $B$ are 1-forms. This other equivalent master action appears to be better suited for the inverse of our problem, that is, of finding the vector picture if the scalar one is already known.

	Resuming the analysis of (\ref{amt}): from its variation with respect to $\phi$, we obtain $dF=0$, which implies $F=dA$; inserting this result into $S_{M*}$, $S_{A*}$ is obtained.	To settle the other side of duality, the variation in regard to $F$ is evaluated, leading to a nontrivial NC extension of Eq.(\ref{map2}) without source, namely,
\be
	\label{phiF}
	\frac 1 {2a} d \phi =   \fdu \( 1 + \la \theta , F \ra \) + \frac 12 \la F,  F \ra \, \tdu.
\ee
In above, the property $ F \wedge \fdu \, \la F, \theta \ra = \la F, F \ra \; \tdu \wedge F$ was employed. Regarding the fields $\vec D$ and $H$, this reads,
\be
	- \frac 1{2a} \vec \nabla \times \phi = \vec D,
\ee
\be
	- \frac 1 {2a} \dot \phi = H.
\ee

To first order in $\theta$, the inverse of the above relations reads

\be
	\label{Fphi}
	\fdu = \frac 1 {2a} d \phi \( 1 - \frac 1 {2a} \la \ddu \phi, \theta \ra \) - \frac 1  {8a^2} \la d \phi, d \phi \ra \, \tdu ,
\ee

\be
	\vec E = - \frac 1 {2a} \( 1 - \frac 1 {2a} \vec \theta \sdot \vec \nabla \times \phi - \frac 1 {2a} \dot \phi \theta \) \, \vec \nabla \times \phi + \frac 1 {8a^2} (\vec \nabla \phi \sdot \vec \nabla \phi - \dot \phi^2 ) \, \vec \theta,
\ee

\be
	B = - \frac {\dot \phi}{2a} \( 1 - \frac 1 {2a} \vec \theta \sdot \vec \nabla \times \phi - \frac 1 {2a} \dot \phi \theta \) - \frac 1 {8a^2} (\vec \nabla \phi \sdot \vec \nabla \phi - \dot \phi^2 ) \, \theta.
\ee

The insertion of the $F$ expression into $S_{M_\theta}$ leads to a NC extension of the scalar field action, namely,
\be
	\label{phi*}
	S_{M_\theta} \lra - \frac {1}{4a}\int d \phi \wedge \ddu \phi \, \( 1 - \frac 1 {2a} \la \ddu \phi, \theta \ra \) = S_{\phi_\theta}.
\ee
	
The correspondence of the equations of motion between vector and scalar models, as expected, is given by $F=dA$ together with Eq.(\ref{phiF}) (and its inverse). Indeed, if $d$ is applied on both sides of Eq.(\ref{phiF}), with $F=dA$, the equation of motion of $S_{A_\theta}$ is obtained; while the application of $d^\star$ on Eq.(\ref{Fphi}) produces the equation of motion of $S_{\phi_\theta}$.

It is straightforward to verify that the map (\ref{phiF}) correctly relates the Hamiltonians and brackets of both representations.

\vspace{.3in}

With the last result, we defined a new scalar field model whose action is, to leading order in the noncommutativity parameter, classically equivalent to the $U(1)$ model of electromagnetic theory in 3D space-time. Although there are cubic terms in the Lagrangian, this duality also holds in the Feynman path integral \footnote{This is just an additional observation, in this work we do not aim to directly deal with quantization issues of NC theories.}. An analogous claim was done in Ref.\cite{ganor} and explicit computation with path integral for the NC extension of the duality of Maxwell-Chern-Simons and self-dual models was done in Ref.\cite{hr}, which presents the same resulting duality of Ref.\cite{nosso}, which does not use the partition function approach. This result can be generalized. Schematically, let $\cl_1 (A)$ and $\cl_2(B)$ be two classically equivalent Lagrangians that are related by the master Lagrangian $\cl_m(A,B)$ whose partition function is 
\be
	Z = \int \cd A \; \cd B \; \exp \[ -i \int \( a_1 A^2 + \theta A^3 + a_2 B A + f(B) \) d^Dx \].
\ee

Integration on $A$ can be converted in a Gaussian integration by introducing two more fields, as follows,
\ba
	 Z & = &  \int \cd A \; \cd B \; \cd C  \; \cd D \; \exp \[ -i \int \( a_1 A^2 + \theta ACC + D(C-A) + \right. \right . \nonumber \\
	&& \left. \left. + \; a_2 B A + f(B) \) d^Dx \].
\ea

Now integration over $A$ can be readily computed, we should replace $A$ by $\frac 1 {2 a_1} (- \theta CC + D - a_2B)$.  Hence, in the above theory, if classical action duality holds for any $\theta$ and partition function duality holds for $\theta = 0$, partition function duality also holds for $\theta \not= 0$. The same arguments are valid to the NC scalar/vector duality here presented.

\vspace{.5in}

\section{Higher $\theta$ order duality}

To second order in $\theta$, (\ref{a*}) reads\footnote{Note that Ref.\cite{ganor} uses a different convention in the differential forms constant factors.} \cite{second, ganor, dayi},
\be
	\label{a*2}
	S_{A_\theta} = \frac a2 \int \[F^{\mu \nu} F_{\mu \nu} \( 1 + \frac 12 \theta^{\mu \nu} F_{\mu \nu} \) + L_{\theta^2} \] d^3x,
\ee
with
\ba
	L_{\theta^2} =&& -2 \; \tr (\theta F \theta F^3 ) + \tr (\theta F^2 \theta F^2) + \tr (\theta F) \; \tr(\theta F^3) - \frac 18 \tr (\theta F)^2 \; \tr (F^2) + \nonumber \\
	&& + \frac 14 \tr(\theta F \theta F) \; \tr (F^2)
\ea
and $\tr (AB) = A_{\mu \nu} B^{\nu \mu}$, $\tr (ABCD) = A_{\mu \nu} B^{\nu \lambda} C_{\lambda \kappa} D^{\kappa \mu} \;$ \emph{etc}.

Fortunately, in the 3D space-time, the above expression can be considerably simplified. We have already used in Eq. (\ref{a*}) that $\tr (FF \theta F) = \frac 12  \tr (FF) \; \tr (F \theta)$, with some reflection this relation can be generalized to 
\be
	\label{rel}
	\tr (A B_1 A B_2 \; ... \; A B_n) = \( \frac 12 \)^{n-1} \prod^n_{k=1} \tr (A B_k),
\ee
for any anti-symmetric 2-rank tensors $A, \{ B_k \}$. Therefore,
\be
	\label{ls}
	L_{\theta^2} = \frac 14 \tr (FF) \; \tr( \theta F)^2.
\ee

The master action $S_{M_\theta}$ (\ref{amt}) can now be extended to second order in $\theta$, this is achieved by adding $- a \int {^\star F} \wedge F \; \la F, \theta \ra^2$ to the first order expression. Thus,

\ba
	^\star F = && \frac {d \phi} {2a} \( 1 - \frac {\la \theta, {^\star d \phi} \ra}{2a} - 3 \frac {\la \theta, { ^\star d \phi} \ra^2}{ 4a^2} + \la \theta, \theta \ra \frac {\la d \phi, d \phi \ra }{ 8a^2} \) - \nonumber \\
	\label{fm}
	&& - {^\star} \theta \frac {\la d \phi,  d \phi \ra }{8 a^2} \( 1 -  5 \frac {\la \theta, {^\star} d \phi \ra} {2a} \)
\ea
and
\be
	\label{p*2}
	S_{\phi_\theta} = - \frac 1 {4a} \int d \phi \wedge {^\star} d \phi \( 1 - \la  \tht, d \phi \ra + 3 \la  \tht, d \phi \ra^2 + \frac 14 \la \tht, \tht \ra \; \la d \phi, d \phi \ra \),
\ee
where $\tht = {^\star} \theta / 2a$. Hence, in the scalar picture, at least to second order, $\tht$ is the Lorentz violation parameter and $\theta$ is unnecessary. Note that only through the employment of $\tilde \theta$ the coupling constant $a$ of the original gauge theory appears in the dual picture as a global factor $a^{-1}$. \emph{A priori}, one can even conjecture that $\tht$ is the fundamental parameter of the scalar picture, while $\theta$ is inferred by duality. Nevertheless, unless the slowly varying fields limit is employed, this is just an illusion of a non-exact symmetry. 

Starting from the third order expansion in $\theta$, terms with more derivatives than potentials appear in the Seiberg-Witten map of $\hat F$ and are present in $L_{\theta^3}$, as we will show (any $L_{\theta^n}$ can only depend on $A$ through $F$, but it can have more derivatives than $A$'s). These factors spoil the last suggested symmetry. To infer these terms, we will use the following Seiberg-Witten differential equation \cite{sw}
\ba
	 \label{dsw}
	 \delta \hat \fmnd	(\theta) &=& \frac 14 \delta \theta^{\alpha \beta} \[ 2 \hat F_{\mu \alpha} * \hat F_{\nu \beta} + 	2 \hat  F_{\nu \beta} * \hat F_{\mu \alpha}  - \hat A_\alpha * (\hat D_\beta \hat F_{\mu \nu} + \partial_\beta \hat F_{\mu \nu})  - \right. \\ \nonumber
	&&  \left. - (\hat D_\beta \hat F_{\mu \nu} + 	\partial_\beta \hat F_{\mu \nu}) * \hat A_\alpha \].
\ea

Expanding $\hat F$ and $\hat A$ in powers of $\theta$, to third order it reads
\be
	 \label{dsw3}
	 \delta \hat \fmnd^{(3)}	(\theta) = - \frac 14 \delta \theta^{\alpha \beta} \theta^{\alpha' \beta'}\theta^{\alpha'' \beta''} \( \prt_{\alpha'}\prt_{\alpha''}F_{\mu \alpha} \prt_{\beta'}\prt_{\beta''}F_{\nu \beta} - \prt_{\alpha'}\prt_{\alpha''} A _\alpha \prt_{\beta'}\prt_{\beta''} \prt_\beta F_{\mu \nu} \) +...
\ee
Where $F_{\mu \nu} = \hat F_{\mu \nu}^{(0)}$ and $A_\mu = A^{(0)}_\mu$. Only the terms with more derivatives than fields were written in the above expression. Inserting this result into Eq.(\ref{sm}), the only terms of $L_{\theta^3}$ which have more derivatives than fields are in the following expression\footnote{This solution can also be inferred by the results of Ref.\cite{fidanza}, Section 3.2, in which the Seiberg-Witten map is expanded in powers of $A$.}
\be
	\label{df}
	\theta^{\alpha \beta} \theta^{\alpha' \beta'} \tr(\prt_{\alpha} \prt_{\alpha'} F \; \theta \; \prt_{\beta} \prt_{\beta'} F \; F) - \frac 14 \theta^{\alpha \beta} \theta^{\alpha' \beta'} \; \tr(F \theta) \; \tr (\prt_{\alpha} \prt_{\alpha'} F \; \prt_{\beta} \prt_{\beta'} F). 
\ee

The contribution of these terms to the equations of motion is given by 
\be
	\label{ec}
	\theta^{\alpha \beta} \theta^{\alpha' \beta'} \prt_\mu \[ F_{\alpha \alpha'}^\mu \; \theta \; F_{\beta \beta'}^\kappa + \frac 12 \tr(F_{\alpha \alpha'} \theta) \; F^{\kappa \mu}_{\beta \beta'} +   F_{\alpha \alpha'}^{[\mu}  \; F_{\beta \beta'} \; \theta^{\kappa]} + \frac 14 \tr(F_{\alpha \alpha'} F_{\beta \beta'}) \; \theta^{\kappa \mu} \].
\ee
In above, we introduced a compact notation: non-explicit indices are contracted like in matrices, extra indices in $F$ are derivatives and $F_{\alpha \alpha'}^{[\mu}  \; F_{\beta \beta'} \; \theta^{\kappa]} = F_{\alpha \alpha'}^{\mu}  \; F_{\beta \beta'} \; \theta^{\kappa} -  F_{\alpha \alpha'}^{\kappa}  \; F_{\beta \beta'} \; \theta^{\mu}$. For instance, the first term in (\ref{ec}) reads $\prt_\alpha \prt_{\alpha'} F^\mu_{\;\; \nu} \; \theta^{\nu \lambda} \; \prt_\beta \prt_{\beta'} F_\lambda^{\;\; \kappa}$.

A careful analysis of the symmetries and anti-symmetries of each term of (\ref{ec}) and their linear independence for arbitrary $\theta$ and $D \ge 4$ shows that (\ref{ec}) is not null. To directly assure unambiguously in any dimension ($D \ge 2$) that (\ref{ec}) is not the trivial identity (or that (\ref{df}) is not a surface term or null) one may evaluate a particular case of (\ref{ec}); for instance, for $D\ge 3$, let $\kappa = 2$ and $\theta$ be equal to zero except for the components $\theta^{01}$ and $\theta^{10}$, namely,
\ba
	&& \( \theta^{10} \)^3 \[ \prt_\mu \( \stackrel{..}{F}^{\mu \, [0} F''^{\, 1]\, 2} + F''^{\, \mu \, [0} \stackrel{..}F^{\, 1]\, 2} - 2 \, \dot F'^{\mu \, [0} \dot F'^{\, 1] \,2} +  \stackrel{..}{F}^{\, 1 0} F''^{\, 2 \mu} + \right. \right.  \\  
	&& \left. \left. + F''^{\, 1 0} \stackrel{..}F^{\, 2 \mu} - 2 \, \dot F'^{\, 1 0} \dot F'^{\, 2 \mu}\) + \prt_{[0} \( \stackrel{..}{F}^{\, 2 \nu} F''_{\, 1] \, \nu} + F''^{\, 2 \nu} \stackrel{..}F_{\, 1] \, \nu} - 2 \, \dot F'^{\, 2 \nu} \dot F'_{\, 1] \, \nu} \) \], \nonumber 
\ea
where each dot and each prime means, respectively, $\prt_0$ and $\prt_1$. The above expression is not identically null in any dimension (greater than two). This result is in conflict with a certain proposition of Ref.\cite{bc}, see our Conclusions for more details.

The expressions (\ref{dsw} - \ref{ec}) are valid for arbitrary space-time dimensions. Once again, in the 3D space-time a considerable simplification is possible. Although the property (\ref{rel}), in that form, cannot be used in (\ref{df}), a straightforward computation shows that an analogous result is valid. In the 3D space-time, the expression (\ref{df}) is equal to
\be
	\label{df3d}
	\frac 14 \; \theta^{\alpha \beta} \theta^{\alpha' \beta'} \; \tr( F_{\alpha \alpha'} F_{\beta \beta'}) \; \tr(F \theta).
\ee 

Adhering to the third order expansion, the contribution of the above expression to $S_{\phi_\theta}$ (\ref{p*2}) is obtained by the replacement $F \rightarrow {^\star} d \phi / (2a)$. Consequently, to third order in $\theta$, $S_{\phi_\theta}$ cannot be expressed only through $\tilde \theta$, $\theta$ is also necessary\footnote{One may artificially insert $\ep$'s in order to change $\theta^{\alpha \beta} \prt_\alpha \prt_\beta$ to $\propto \tilde \theta_{\mu} \ep^{\mu \alpha \beta} \prt_\alpha \prt_\beta$, this procedure is innocuous since ${^\star} \tilde \theta \propto \theta$; but we are adopting the rule to always write or $\theta$, or $\tilde \theta$, never ${^\star} \tilde \theta$. Moreover this procedure does not avoid the difficulties with $S$-duality, since $\tilde \theta$ will not occur proportionally to $\phi$ in the dual picture.}. This violates the symmetry between $\theta$ and $\tilde \theta$ present in electromagnetic duality up to the second order in $\theta$. Consequently, in the scalar picture, the constant $a$ does not appear as a global $a^{-1}$ and $S$-duality is broken (at least in regard to its usual form).

In the slowly varying fields limit, the terms in the Seiberg-Witten mapped action which depend on the derivatives of $F$ are neglected, therefore $S_{\phi_\theta}$ to third order in $\theta$ can be solely expressed in terms of $\tilde \theta$. In this limit, since the Seiberg-Witten mapped Lagrangian is a function of $F$ alone  (without derivatives) \cite{sw}, the Lagrangian is expressed as a function of $\tr (FF)$ and $\tr(F \theta)$ only (due to Eq.(\ref{rel})); therefore the dual scalar action $S_{\phi_\theta}$ to all orders in $\theta$ can be expressed by $\tilde \theta$, without reference to $\theta$ (or ${^\star} \tilde \theta$). Although the property (\ref{rel}) is in general false in the 4D space-time, the dual action can also be expressed by $\tilde \theta$ alone in the 4D space-time, to all orders in $\theta$, if the slowly varying fields limit is used \cite{aschieri}. The relation (\ref{rel}) simplifies considerably the work in the 3D analysis. 

\vspace{.5in}

\section{Conclusions}

In this Letter we establish, to second order in $\theta$, the scalar description of 3D NC electromagnetic theory, which is usually described by the gauge model in Eq.(\ref{sm}). We show that the rule $\theta \rightarrow \tilde \theta = {^\star \theta g^2}$, found in Ref.\cite{ganor} in the context of 4D NC electromagnetic duality, can be extended to the 3D case up to second order in $\theta$, Eqs.(\ref{a*2})(\ref{p*2}). With this rescaling of $\theta$, the coupling constant of one model becomes the inverse of the other. This is indeed a curious relation between these dual models, but this relation is only approximately valid: starting from the third order $\theta$ expansion, in general it becomes false in both 3D and 4D cases. The coupling constant does not appear proportionally to $\theta$, but to $\phi$ instead; so, to any order, it is possible to do the replacement $\phi \rightarrow \phi g^2$ and the final answer is a non-inversion of the coupling constant. In the 4D case, \emph{a priori} it is possible to think that somehow the coupling constant appears proportionally to $\theta$, this doubt however is absent from the 3D case, for $g^2$ is dimensionful in this space and $\theta$ appears proportionally to $\prt \prt$ in some terms, like those in Eq.(\ref{df}). Since, up to the subleading order in $\theta$ or in the slowly varying fields limit, the 4D duality connects two $U(1)$ theories, one with $\theta$ and the other with $\tilde \theta$ \cite{ganor, aschieri, 4D}, it might appear that $\theta$ and $\tilde \theta$ could be used indistinguishably; however a simple analysis of the 3D case shows this does not proceed. The 3D clearly states that, if a theory has $\theta$ as its parameter, there is another equivalent one with a \emph{different definition of the fields} which has the parameter $\tilde \theta$ (this is a direct interpretation of the duality map, e.g., Eq.(\ref{fm})). As a final remark of this duality to second order in $\theta$, it is easy to see from the equations of motion and the interchange between $\theta$ and $\tilde \theta$ that the 3D NC duality preserves spacial isotropy (i.e., if one of the dual models is isotropic, the other also is) and, if a spacial anisotropy is present, duality rotates the preferential direction by $\pi /2$.

Currents can be easily inserted in this duality, along the lines of Sec.2, if it is assumed a $\theta$ non-dependent coupling like $A \wedge {^\star J}$ in the mapped action. Nevertheless, this violates correspondence with the $U_*(1)$ theory, which asserts the coupling $\hat A \wedge_* {^\star} \hat J$, whose map was found in Ref.\cite{banerjee}.

In the Sec.IV, we proved, by means of a straightforward calculation valid in any dimension greater than two, that the Seiberg-Witten mapped Lagrangian of the NC electromagnetic theory ($L_{A_\theta}$) depends on $F$ and its derivatives\footnote{Although it was not explicitly shown in Sec.4, it is not hard to evaluate that the particularization of (\ref{df}) to $D=2$ is also different from zero.}. Up to the  second order in $\theta$, the derivatives on $F$ can be combined with the fields $A$ to produce another $F$ (eliminating all the explicit dependence on the $A$'s, since there are the same number of derivatives than $A$'s). Nevertheless, the Seiberg-Witten differential equation (\ref{dsw}) leads to the appearance of terms with more derivatives than fields in the third order expansion. These terms were applied into the NC electromagnetic Lagrangian ($L_{\hat A_*})$ and the resulting terms were stated in (\ref{df}). Perhaps surprisingly, these terms are not null nor are surface terms, as subsequently we have shown\footnote{In general the Seiberg-Witten map is not unique\cite{asakawa}, nevertheless the additional terms do not influence this analysis.}.  This result is not in agreement with the first part of a proposition in Ref.\cite{bc}. We think our result should be considered as a counter-example to it. Indeed, the first part of Proposition 3.1 does not seem to be correct in general \cite{comment}. However, it should be stressed that it clearly holds in the slowly varying fields limit and, in this limit, it is compatible with our results; moreover, any results which depend on that proposition are perfectly valid in that limit. There are some others interesting consequences which we are now evaluating \cite{dm}. 

As previously stated, this work does not aim to resolve string $S$-duality issues in the presence of a magnetic background, like Ref. \cite{stringd} does. However, a certain exchange of $\theta$ with ${^\star} \theta g^2$, among other similarities, occurs in both cases. According with our result, this exchange only occurs to all orders in $\theta$ in the slowly varying fields limit. At the moment it is not clear to us if our result has consequences to the string  $S$-duality of NC theories since, among other possibilities, we may have come across a pathological feature of the Seiberg-Witten map \cite{dm}.

We think further developments of the NC electromagnetic duality can prove useful to construct effective models and to understand NC theories in general.

\acknowledgments

This work is partially supported by PRONEX/CNPq. We thank Prof. R. Banerjee and Prof. V. Rivelles for useful comments and Prof. S.L. Cacciatori and collaborators for their attention to our problem and their cordial answer. DCR also thanks FAPERJ (Brazilian research agency) for financial support.

\end{document}